\documentclass[12pt]{article}
\begin{document}

\title{Bifurcations due to small time-lag in coupled excitable systems}
 \author{Nikola Buri\' c  \thanks{e-mail: buric@phy.bg.ac.yu},
and Dragana Todorovi\' c   \\ Department of Physics and
Mathematics,\\ Faculty of Pharmacy, University of Beograd,\\
Vojvode Stepe 450, Beograd, Serbia and Montenegro . }

\maketitle

\begin{abstract}
A system of ODE's is used to attempt an approximation of the
dynamics of two delayed coupled FitzHugh-Nagumo excitable units,
described by delay-differential equations. It is shown that the
codimension 2 generalized Hopf bifurcation acts as the organizing
center for the dynamics of ODE's for small time-lags. Furthermore,
this is used to explain important qualitative properties of the
exact dynamics for small time-delays.

\end{abstract}

\newpage

{\it Introduction}

\vskip 0.5cm
 Dynamics  of a pair of excitable systems with
time-delayed coupling is quite different from the dynamics with
the instantaneous coupling. Also, the behaviour of coupled
excitable systems differs from that of coupled oscillators with
the same coupling. A prime example of the type II  excitable
behaviour (see [Izhikevich, 2000]) has been the system introduced
by FitzHugh [FitzHugh 1955] and Nagumo et all [Nagumo et all,
1962], as an approximation of the Hodgine-Haxley  model of the
nerve cell membrane. One form of the FHN equations is (see
[Murray, 1992]):
\begin{eqnarray}
\dot x & = & -x^3 + (a + 1)x^2 - a x -y+I, \nonumber \\
 \dot y & = & b x -\gamma y
\end{eqnarray}
As is well known the system (1) can operate in two regimes
depending on the
 external current $I$, as an excitable system if $I=0$ and
 \begin{equation}
 4{b\over \gamma}<(a-1)^2,
 \end{equation} with the
 stable stationary solution as the only attractor, or as a
 relaxation oscillator $I=const\neq 0$, with the
 stable limit cycle as the only attractor.

Two delayed coupled FHN excitable systems with delayed coupling
given by the following equations:
\begin{eqnarray}
 \dot x_1 & = & -x_1^3 + (a +
1)x_1^2 - ax_1 -y_1+c\tan^{-1}(x^{\tau}_{2}),\nonumber \\ \dot y_1
& = & bx_1 - \gamma y_1,\nonumber \\ \dot x_2 & = & -x_2^3 + (a +
1)x_2^2 - ax_2 -y_2+c\tan^{-1}(x^{\tau}_{1}),\nonumber \\ \dot y_2
& = & bx_2 - \gamma y_2,
\end{eqnarray}
where $x^{\tau}(t)=x(t-\tau)$, have been recently analyzed in
[Buri\'c$\&$Todorovi\'c, 2003]. Motivation, background and the
relevant literature is discussed in detail in
[Buri\'c$\&$Todorovi\'c,  2003] and will not be repeated here.

Besides different types of oscillations induced by coupling and
changed by the time-delay, the system displays two different types
of excitable behaviour. The first one is described by a single
stationary solution at $E_0\equiv(0,0,0,0)$ as the only attractor.
The other type is described by two coexisting attractors, the
stable stationary solution $E_0$ and a stable limit cycle
corresponding to periodic exactly synchronous excitations of the
two units. The later regime occurs only in a specific, relatively
small, domain in the parameters $(c,\tau)$ plane.
Buri\'c$\&$Todorovi\'c obtained bifurcation curves in $(c,\tau)$
plane by solving the characteristic equation of (3) for the
non-hyperbolic roots. The bifurcation curves served as a guide for
the analyzes of the system by extensive numerical calculations,
which resulted in the classification of possible excitable and
oscillatory dynamics.

In this letter we would like to report some analytic results aimed
at better understanding of the types of bifurcations which are
relevant for the two types of the excitable behaviour. To this end
we shall analyze the bifurcations of the stationary solution that
occur in a system related to the equations (3) in the following
way. Both types of excitable behaviour happen for relatively small
time-lags, and such a small time-lag induces the responsible
bifurcations. Thus it might be justified to replace the
time-delayed argument of the coupling function in (3) by the
following approximation:
\begin {equation}
f(x(t-\tau))\approx f(x-\tau \dot x).
\end{equation}
and approximate the delay-differential (DDE)  by ordinary
differential equations. The approximate system is given by:
\begin{eqnarray}
\dot x_1 & = & -x_1^3 + (a + 1)x_1^2 - ax_1 -y_1+\nonumber\\&+&c
\tan^{-1}(x_2-\tau (-x_2^3 + (a + 1)x_2^2 - ax_2
-y_2+c\tan^{-1}(x_1))),\nonumber
\\ \dot y_1 & = & bx_1 - \gamma y_1,\nonumber \\ \dot x_2 & = &
-x_2^3 + (a + 1)x_2^2 - ax_2
-y_2+\nonumber\\&+&c\tan^{-1}(x_1-\tau ( -x_1^3 + (a + 1)x_1^2 -
ax_1 -y_1+c\tan^{-1}(x_2))),\nonumber
\\ \dot y_2 & = & bx_2 - \gamma y_2,
\end{eqnarray}

  Validity of the approximation (4) and (5) is not {\it apriori} justified even for small $\tau$.
 Such type of questions have been
 analyzed before in general and for specific examples (see [Meinardus $\&$ Nuenberg, 1985]). More
 recent example of such an analyzes, for a particular system, is given in the reference
  [Faro $\&$ Velasco, 1997], where the
 approximation has been investigated using a predator-prey
 equations with a time-delay by comparison of
   numerically obtained bifurcation curves.

The main result of our analyzes is that the bifurcation which acts
as the organizing center for the dynamics of the system (5) for
small $\tau$ is the codimension 2 generalized Hopf (Bautin)
bifurcation. Furthermore, the bifurcation occurs for quite a small
time-lag where the bifurcation curves of the exact and approximate
system almost coincide and the dynamics is qualitatively the same.
This fact is used to explain the occurrence of the two types of
excitable behaviour.

The approximate and the exact systems will be abbreviated as $\dot
X={\cal F}(X,X^{\tau})$ and $\dot X={\cal F}_{app}(X)$ where $X\in
{\bf R}^4$ represents the collection of coordinates
$x_1,y_1,x_2,y_2$.

\vskip 1cm
 {\it Bifurcations of the stationary solution}
\vskip 0.5cm

We shall restrict attention only on such values of the parameters
$a,b,\gamma,c$ that the system ${\cal F}_{app}$ for $\tau=0$ has
only one stationary solution. This occurs if
\begin{equation}
c<c_1\equiv a+b/\gamma.
\end{equation}
Futhermore, each unit is excitable when decoupled, i.e. the
condition (2) is assumed satisfied.
 We fix the parameters $a,b$ and $\gamma$ to some
arbitrary such value,
 and consider the bifurcations of $E_0$
that could occur as the parameters $c$ and $\tau$ are varied. The
bifurcation set ${\cal B}_{E_0}$ is defined as the set of
$(c,\tau)$ values such that the stationary solution $E_0$ is not
hyperbolic. By an abuse of notation, we shall use the same symbol
${\cal B}_{E_0}$ for the part of ${\cal B}_{E_0}\subset {\bf
R}^+\times {\bf R}^+$ satisfying (6), i.e.
\begin{equation}
{\cal B}_{E_0}=\left\{ (c,\tau)\in {\bf R}^+\times {\bf R}^+| {\bf
Re}\lambda(c,\tau)=0, c<c_1\right\},
\end{equation}
 where $\lambda$ is any root of the characteristic polynomial.

The  set ${\cal B}_{E_0}$ consists of two line segments in
$(c,\tau)$ plane. Namely:
\begin{eqnarray}
{\cal B}_{E_0}&=&\left\{ (c,\tau)| \tau=1/c, c<c_1\right\}\bigcup
\left\{ (c,\tau)| \tau={c-a-\gamma\over c(c-a)}, c\in
(a,c_1)\right\}\nonumber\\ &\equiv &{\cal B}_{E_0;p}\bigcup{\cal
B}_{E_0;H}.
\end{eqnarray}

Indeed, the linear part of (5)
\begin{equation}
A=\pmatrix{F&-1&D&E\cr b&-\gamma&0&0\cr D&E&F&-1\cr
0&0&b&-\gamma},
\end{equation}
where
\begin{equation}
F=-a-c^2\tau,\quad D=c+ca\tau,\quad E=c\tau
\end{equation}
 implies the following characteristic equation:
 \begin{eqnarray}
 \Delta(\lambda)&\equiv&\Delta_1(\lambda)\Delta_2(\lambda)=0
 \end{eqnarray}
 where
 \begin{eqnarray}
 \nonumber\\
\Delta_1(\lambda)&=&\lambda^2+(\gamma-F-D)\lambda+b-F\gamma-\gamma
D-bE\nonumber\\ \Delta_2(\lambda)&=&
\lambda^2+(\gamma-F-D)\lambda+b-F\gamma+\gamma D+bE
\end{eqnarray}
 with solutions
\begin{eqnarray}
\lambda_{1,2}=\left[-\gamma+F+D\pm\sqrt{(\gamma+F+D)^2-4b(1-E)}\right]/2,\>
\Delta_1(\lambda_{1,2})=0, \nonumber\\
\lambda_{3,4}=\left[-\gamma+F-D\pm\sqrt{(\gamma+F-D)^2-4b(1+E)}\right]/2,\Delta_2(\lambda_{3,4})=0.
\end{eqnarray}
 A nonhyperbolic root can be either equal to zero or pure
imaginary. In the first case,
\begin{eqnarray}
\Delta_1(0)=0& \Leftrightarrow & b-F\gamma-g D-b E=0
 \nonumber\\ &\Leftrightarrow & \tau c (c\gamma-a \gamma-b)=c\gamma-a \gamma-b,
 \end{eqnarray}
 which defines the line segment ${\cal
B}_{E_0;p}$. The second factor of the characteristic polynomial
has no zero roots for any
 positive $c$ and $\tau$. In the second case
\begin{eqnarray}
\Delta_1(iv)=0,\> v>0 & \Leftrightarrow
&-v^2+(\gamma-F-D)iv+b-F\gamma-\gamma D-b E =0
 \nonumber\\ &\Leftrightarrow & v^2=b-F\gamma-\gamma D-b E>0\>{\rm and }\>
 (\gamma-F-D)v=0.\nonumber
 \end{eqnarray}
If $c\in (a,c_1)$ then from the last condition we obtain the line
segment ${\cal B}_{E_0;H}$ and
$v=\sqrt{\gamma(a\gamma+b-c\gamma)/(c-a)}$. On the other hand, if
$c \notin (a,c_1)$ there is no pure imaginary solution of
$\Delta_1=0$. Furthermore $\Delta_2=0$ has no pure imaginary
solutions for any positive $\tau$ and $c$. Thus, ${\cal B}_{E_0}$
is indeed given by (8). It is illustrated in figure 1a. The point
where ${\cal B}_{E_0}$ intersects the $c$ axis is denoted by
$c_0$. Thus:
\begin{equation}
c_0=a+\gamma.
\end{equation}

The type of bifurcations occurring for the parameters $(c,\tau)$
in  ${\cal B}_{E_0;p}$ and  ${\cal B}_{E_0;H}$ are described in
the following two theorems.

{\bf Theorem 1} The system ${\cal F}_{app}$ has a pitchfork
bifurcation for any $(c,\tau)\in  {\cal B}_{E_0;p}$.

Proof: The linear part $A$ on ${\cal B}_{E_0;p}$ has a simple zero
eigenvalue. The type of bifurcation at $\tau=1/c$ is analyzed by
reducing the system on the corresponding  center manifold with the
 parameter $\epsilon=\tau-1/c$. As we shall see, it is enough to
 consider the extended system, given by (5) and $\dot\epsilon=0$
 expended up to the third order in $x_1,x_2,y_1,y_2,\epsilon$:
 \begin{eqnarray}
 \dot x_1&=&-(a+c)x_1-y_1+(a+c)x_2+y_2+(a+1)(x_1^2-x_2^2)\nonumber\\
&+&(c/3-1)(x_1^3-x_2^3)-ax_2^3-x_2^2y_2+cx_2^2x_1-c(a+1)\epsilon
x_2^2+\nonumber\\ &+&ac\epsilon x_2+c\epsilon y_2-c^2\epsilon
x_1\nonumber\\ \dot y_1&=&bx_1-\gamma y_1\nonumber\\
 \dot x_2&=&(a+c)x_1+y_1-(a+c)x_2-y_2-(a+1)(x_1^2-x_2^2)\nonumber\\
&-&(c/3-1)(x_1^3-x_2^3)-ax_1^3-x_1^2y_1+cx_1^2x_2-c(a+1)\epsilon
x_1^2+\nonumber\\ &+&ac\epsilon x_1+c\epsilon y_1-c^2\epsilon
x_2\nonumber\\ \dot y_2&=&bx_2-\gamma y_2\nonumber\\ \dot
\epsilon&=&0.
\end{eqnarray}

 The center manifold with the parameter $\epsilon$ of the
system (16), in the new coordinates $(x,y,z,t,\epsilon)$ related
to the old ones by
\begin{eqnarray}
x_1 &=&x-z-t ,\nonumber\\
 y_1 &=& bx/\gamma+y-bz/(\gamma+\lambda_3)-bt/(\gamma+\lambda_4),\nonumber\\
  x_2 &=& x+z+t,\nonumber\\
  y_2&=&  bx/\gamma+y+bz/(\gamma+\lambda_3)+bt/(\gamma+\lambda_4),\nonumber\\
  \lambda_{3,4}&=&(-2(a+c)-\gamma\pm\sqrt{(2a+2c-\gamma)^2-8b})/2,
\end{eqnarray}
is:
\begin{eqnarray}
W^c(0)=\{(x,y,z,t,\epsilon)|y&=&h_1(x,\epsilon),z=h_2(x,\epsilon),t=h_3(x,\epsilon);\nonumber\\
&&h_i(0,0)=0,\> Dh_i(0,0)=0,\> i=1,2,3 \},
\end{eqnarray}
where
\begin{eqnarray}
 h_1(x,\epsilon)&=&-{bc\over
\gamma^2}(a+b/\gamma-c)x\epsilon+{b\over
\gamma^2}(a+b/\gamma-c)x^3+{bc\over
\gamma}(a+1)x^2\epsilon+\dots,\nonumber\\
h_2(x,\epsilon)&=&0,\nonumber\\ h_3(x,\epsilon)&=&0 .
\end{eqnarray}
 Restriction of (16) on the center manifold (18) is given by:
\begin{equation}
\dot x\equiv F(x,\epsilon)=
c(a+b/\gamma-c)x\epsilon-(a+b/\gamma-c)x^3-c(a+1)x^2\epsilon+\dots,
\end{equation}
and satisfies:
 $$ {\partial F(0,0)\over \partial \epsilon}=0,\quad
{\partial^2 F(0,0)\over \partial \epsilon^2}=0,
$$
 $$ {\partial^2
F(0,0)\over \partial \epsilon\partial x}=c(a+b/\gamma-c)\neq 0, $$
$$ {\partial^3 F(0,0)\over \partial x^3}=-6(a+b/\gamma-c)\neq 0.
 $$
These are the sufficient and necessary  conditions for the
pitchfork bifurcation.
  The system (5), under the
condition (6) is, in a neighborhood of $(x,\epsilon)=(0,0)$
locally topologically equivalent to $ \dot x=\epsilon x-x^3 $ (see
[Arrowsmith, 1990], [ Kuznetsov, 1995]). Thus, if (6) is satisfied
then for $\tau\sim 1/c$, and $\tau < 1/c$ the stationary solution
$E_0$ is stable. For $\tau
> 1/c$ the stationary point $E_0$ is unstable but there are two
new stable stationary solutions. $\clubsuit$

 {\bf Theorem 2} For  the parameter values $(c,\tau)\in {\cal
B}_{E_0;H}$ the system ${\cal F}_{app}$ has either the
supercritical Hopf or the subcritical Hopf or  the generalized
Hoph bifurcation. Furthermore, there are such values of $a,b$ and
$\gamma$ that the value $c_B$ for which the system has the
generalized Hopf bifurcation satisfies $c_B\in (c_0,c_1)$.

Proof: For the parameters in ${\cal B}_{E_0,H}$ the matrix $A$ has
a pair of purely imaginary eigenvalues $\lambda_{1,2}=\pm iv, \>
v>0$ and no other nonhyperbolic eigenvalues. Furthermore, for
$(c,\tau)\in  {\cal B}_{E_0;H}$
\begin{equation}
d\equiv{d{\bf Re}\lambda_{1,2}\over d\tau}|_{{\cal
B}_{E_0,H}}={1\over 2}{d (-\gamma+F+D)\over d\tau}|_{{\cal
B}_{E_0,H}}=c(a-c)/2<0.
\end{equation}

 Thus, $(c,\tau)\in
{\cal B}_{E_0;H}$  corresponds to the Hopf bifurcation. The type
of the Hopf bifurcation is determined by studying the normal form
of the system on the two-dimensional invariant center manifold. To
obtain the normal form we use the procedure introduced in [Coullet
$\&$ Spiegel 1983], and applied by Kuznetsov [Kuznetsov, 1997] to
obtain the
 relevant coefficients in normal forms of all codimension 1 and 2
 bifurcations of stationary solutions of ODE.

 As we shall see, it is enough to start with the system ${\cal
F}_{app}$ expanded up to the terms of the third order.
 $$\dot X= A X+{1\over 2}{\cal F}_{app,2}(X,X)+{1\over 6}{\cal
 F}_{app,3}(X,X,X),$$
 where
 $$
{\cal F}_{app,2}(X,X)=\pmatrix{(a+1)x_1^2-c(a+1)\tau x_2^2\cr 0\cr
(a+1)x_2^2-c(a+1)\tau x_1^2\cr 0}, $$ and
 $$
{\cal F}_{app,3}(X,X,X)=\pmatrix{
(c^2\tau/3-1)x_2^2+(c\tau-c/3-ca\tau)x_1^3-c\tau x_1^2y_1+c^2\tau
x_2x_1^2 \cr 0\cr (c^2\tau/3-1)x_2^2+(c\tau-c/3-ca\tau)x_1^3-c\tau
x_1^2y_1+c^2\tau x_2x_1^2\cr 0} $$

First introduce a complex eigenvector $Q\in {\mathbf R}^4$ of $A$,
i.e. $ AQ=iv Q$ with components
 $$Q=\pmatrix{ 1\cr (c-a)(1-iv/\gamma)\cr 1\cr (c-a)(1-iv/\gamma) }$$
and the corresponding eigenvector of $A^T$: $\qquad A^TP=-iv P$
$$P=\pmatrix{ {v+i\gamma\over 4v}\cr {-i\gamma\over 4v(c-a)}\cr
{v+i\gamma\over 4v}\cr {-i\gamma\over 4v(c-a)}}$$ normalized to
$<P,Q>=\bar P^TQ=1$. Vectors $Q$ and $\bar Q$ (the complex
conjugate of $Q$) form a basis in the center-subspace $E^c$ of
$A$, so any vector $R\in E^c$ can be written as $R=\alpha
Q+\bar\alpha \bar Q$ where $\alpha=<P,R>\in {\mathbf{C}}^1$. The
relation between the
 original system $\dot X={\cal F}_{app}$ and the
 the complex normal form of the
system on the center manifold $X=H(\alpha,\bar\alpha)$ of the
following form:
\begin{equation}
\dot\alpha=iv\alpha+l_1 \alpha|\alpha|^2+ l_2\alpha|\alpha|^4+
O(|\alpha|^6)
\end{equation}
is contained in the corresponding homological equation: $$
{\partial H\over \partial \alpha}\dot \alpha+{\partial H\over
\partial \bar{\alpha}}\dot {\bar \alpha}={\cal
F}_{app}(H(\alpha,\bar\alpha)) $$ Substituting the Taylor
expansions of the transformation $H$ and the system ${\cal
F}_{app}$ into the homological equation, and collecting the terms
with the same order one gets the coefficients $l_1,l_2\dots$ in
the normal form. The third order coefficient $l_1$ is given by:
\begin{eqnarray}
l_1&=&{1\over 2} {\bf Re}<P,{\cal F}_{app,3}(Q,Q,\bar Q)+{\cal
F}_{app,2}(\bar Q,(2iv I_4-A)^{-1} {\cal
F}_{app,2}(Q,Q))\nonumber\\ &-& 2{\cal F}_{app,2}(Q,A^{-1}{\cal
F}_{app,2}(Q,\bar Q))> \nonumber\\
 &=& {1\over2}\left [ -(c+3){\gamma\over c-a}+c-a-\gamma+{2\gamma^2(a+1)^2\over
 (c-a)(b+a\gamma-c\gamma)}\right ]\nonumber\\
 &=&{\gamma c^3+B(a,b,\gamma)
 c^2+C(a,b,\gamma)c+D(a,b,\gamma)\over 2(a-c)(b+a\gamma-c\gamma)},
\end{eqnarray}
where
\begin{eqnarray}
B(a,b,\gamma)&=&-b-3a\gamma-2\gamma^2\nonumber\\ C(a,b,\gamma) &=&
2ab+3a^2\gamma+2b\gamma-3\gamma^2+3a\gamma^2\nonumber\\
D(a,b,\gamma)&=&-a^2b-a^3\gamma+3b\gamma-ab\gamma-2\gamma^2
-a\gamma^2-3a^2\gamma^2.
\end{eqnarray}

If the third order coefficient $l_1\neq 0$ the system is locally
smoothly orbitally equivalent to the system $$
\dot\alpha=iv\alpha+l_1 \alpha|\alpha|^2. $$ Since everywhere on
${\cal B}_{E_0,H}$ we have  $d<0$ ,
 the values of the parameters $(c,\tau)\in {\cal B}_{E_0,H}$ such that
 $l_1<0$,
 imply the supercitical Hopf bifurcation,
 if  $l_1>0$ the bifurcation is subcritical,
  and if  $l_1=0$ the bifurcation is of the
generalized Hopf type.
 The denominator of $l_1$ is always negative in the considered
 interval $c\in(a,c_1=a+b/\gamma)$.
  The numerator of $l_1$ always has at least one real zero
 $c_B$. Thus, there is at least one value of $c\equiv c_B$ such
 that the bifurcation is of the generalized Hopf type.
 The following choice of $a,b$ and $\gamma$: $a=0.25,b=\gamma=0.02$ is an example of the
  values such that
$c_b=0.289024\in(c_0,c_1)=(0.27,1,27)$. In this case all three
alternatives occur as $c$ is varied in the interval $(c_0,c_1)$.
 $\clubsuit$

 In the case $c_B\in(c_0,c_1)$ all three types of Hopf bifurcation
  occur.
 In general, the  line segment of codimension 1 subcritical Hopf bifurcations,
joins with the line segment of codimension 1 supercritical Hopf
bifurcations  at some point $(c_B,\tau_B)$ of the codimension 2
 generalized Hopf bifurcation. It is important to point out that
 in all numerical test that we have performed for the values of
 $a,b$ and $\gamma$ such that the isolated unit shows excitable
 behaviour, we have observed precisely the situation where
 $c_B\in(c_0,c_1)$ and actually is quite close to $c_0$.

  As is well known from the theory of
 the codimension 2 generalized Hopf bifurcation, besides the two lines of
Hopf bifurcations of the stationary solution, there is also a line
of fold limit cycle bifurcations emanating from the point
$(c_B,\tau_B)$. For the parameter values in between the lines of
subcritical Hopf and fold limit cycle bifurcations the system
${\cal F}_{app}$  has a stable stationary solution surrounded by
an unstable limit cycle, which is surrounded by a stable limit
cycle, all three lying in the manifold  $x_1=x_2, y_1=y_2$. The
generalized Hopf bifurcation at $(c_B,\tau_B)$ is illustrated in
figure 1b.

Finally, let us remark that the theorems 1 and 2 are probably
correct if the $\tan^{-1}$ coupling function is replaced by any
function of the sigmoid form, although we do not gave the proof in
the general case because the algebra gets rather tedious. The only
conditions that should be required are that:
$f(0)=0,f'(0)>0,f''(0)=0$ and $f'''(0)\neq 0$.

\vskip 1cm
 {\it Approximate vs exact system}
\vskip 0.5cm

The bifurcation set in $(c,\tau)$ plane of the exact system ${\cal
F}$ under the same conditions (2) and (6) on the parameters
$a,b,\gamma$ and $c$, was obtained before in [Buri\'c $\&$
Todorovi\'c, 2003]. Using the bifurcation set and the numerical
test, it was conjectured that there is a domain in $(c,\tau)$ that
corresponds to the death of oscillations due to time-delay. On the
bases of the numerical evidence it was also conjectured that the
bifurcation mechanism beyond the oscillator death is more
complicated than commonly found in oscillators coupled by
diffusion with delay, i.e. the inverse supercritical Hopf
bifurcation (see [Reddy et. all., 1999]).

 We would like to
 use the results about the bifurcations of the approximate system
 ${\cal F}_{app}$
in order to discuss the time-delay death of oscillations that have
been induced by coupling of excitable systems. To this end, we
first compare the bifurcation sets in $(c,\tau)$ plane of the
exact end the approximate system. The two sets are illustrated in
figure 2. The curves denoted $\tau_1$ and $\tau_2$ correspond to
the first and second factor of the characteristic equation of the
exact system, and there $d{\bf Re}\lambda_{1,2}/d\tau<0$ on
$\tau_1$ and  $d{\bf Re}\lambda_{3,4}/d\tau<0$ on $\tau_2$, where
$\lambda_{1,2,3,4}$ are the solutions of the exact characteristic
equation.

 Few observations are in order. Firstly, the approximate systems
 badly fails to describe the bifurcations of the
 exact system for most values of the time-lag. Qualitative
 agreement between the dynamics of the two systems is obtained
 only for very small values of the time-lag. The entire family of
  bifurcations due to nonhyperbolicity of the second factor of the
  characteristic equation of the exact system is missing. The smallest
   time-lag for these
  bifurcations to occur, is to large to be capture by the
  approximation, for all values of $c<c_1$. Furthermore, the
  approximate system has a line of pitchfork bifurcations,
  destabilizing $E_0$ and
  introducing two new stable stationary solutions for large $\tau$
  and any $c$. There is no analogous situation in the exact
  system.

   On the other hand, there is a small but important domain of
  $(c,\tau)$ where the bifurcation curves of the exact system are
 well approximated by the approximate one. This is the domain
 precisely around the generalized Hopf bifurcation, and is
  indicated in the figure 2. Thus, we can use theorem 2,
about the bifurcations of the approximate system, to explain the
 dynamics of the exact system near the bifurcation line
 $\tau_1(c)$. This also supports the conjecture that the mechanism involved in the
 death of oscillations due to time-delay in the exact system involves
 the line of subcritical Hopf and the line of fold
 limit cycle bifurcations organized by the generalized Hopf
 bifurcation.  Consider the system for $c>c_B$ and zero or  small
 time-lag, when it consists of two exactly synchronized
 oscillators (compare figures 3 and 4). The only attractor is the limit cycle in $x_1=x_2$,
 $y_1=y_2$ plane. Increasing the time-lag leads to the subcritical
 Hopf bifurcation for $(c,\tau)\in {\cal B}_{E_0,H}$ when the
 stationary solution becomes stable and an unstable limit cycle is
 created in the same plane as the stable limit cycle.
  The system is bi-stable with the stable stationary state and
  periodic excitations described by the
   stable limit cycle. The unstable limit cycle acts as a threshold.
      Further increase of the time lag leads to the disappearance
      of the two limit cycles  in the fold limit cycle
      bifurcation. This corresponds to the death of oscillatory
 excitations. Upon further increase of $\tau$ the approximate
 system hits the line of pitchfork bifurcations, the stationary
 point becomes unstable and there are two new stable stationary
 solutions. Such dynamics does not occur in the
  exact system for any value of the time-lag. The sequence of
  different attractors obtained for fixed $c$ and successively
  larger $\tau$ for the approximate system, illustrated in figure 3,
  corresponds qualitatively to the sequence in the
   exact system illustrated in figure 4.
    Large time-lags, introduce qualitatively different dynamics
(figures 3d and 4d).  The qualitative correspondence between the
exact system and the approximation is lost.

\vskip 1cm
 {\it Summary and conclusions}
\vskip 0.5cm

We have performed an analyzes of bifurcations of the stationary
solution of a model of two coupled FitzHugh-Nagumo excitable
systems. The model ODE's originate as a small time-lag
 approximation of the DDE's which explicitly include the
 time-delay in the transmission of excitations between the units.

 The main results are given in the two theorems 1 and 2. The
 second theorem identifies the degenerate Hopf bifurcation as the
 main organizing center that is enough to explain all qualitative
 features of the dynamics for small time-lags. There are three
 possible types of dynamics in this domain. The system could be
 excitable, with the stable stationary solution as the only
 attractor, or oscillatory, when the limit cycle is the only
 attractor, or, finally, the system could be bi-stable. In the
 last case there are the stable stationary solution and the stable
 limit cycle. These are the three types of the dynamics that have
 been observed also in the exact system of DDE for small
 time-lags.
 The first theorem determines the boundary in the $(c,\tau)$ plain
 beyond which the dynamics of the approximate system is
 qualitatively different from anything that occurs in the exact
 system for the considered parameter values.

Our analyzes is carried out using an explicit coupling function.
However, the results would probably be the same for any coupling
of the same form with the function $f(x)$ satisfying
$f(0)=0,\>f'(0)>0,\>f''(0)=0$ and $f'''(0)\neq 0$. On the other
hand, different type of coupling, for example of the diffusive
form, would result in different bifurcations and dynamics.

There is yet another set of codimension 2 bifurcations that occur
in the exact system for larger time-delays. They happen at the
intersection of Hopf bifurcation curves, and are quite important
for the properties of the oscillatory dynamics at larger
time-lags. This Hopf-Hopf bifurcations are not captured by the
finite dimensional approximation by ODE's. In order to analyze
them an infinite dimensional generalization of the method applied
in theorem 2 (see [Faria $\&$ Magelhes, 1995], [Schayer $\&$
Campbell, 1998]) could be applied.

 \vskip 1cm
{\bf Acknowledgements} This work is supported by the Serbian
Ministry of Science contract No. 101443. \vskip 2cm

\vskip 2cm
{\bf REFERENCES}
\vskip 1cm

{Izhikevich E.M. [2000] Neural Excitability, Spiking and Bursting.
Int. J. Bif. Chaos, {\bf 10}: 1171-1266.}

{ FitzhHugh R [1955] Mathematical Models of Threshold Phenomena in
the Nerve Membrane. Bulletin of Mathematical Biophysic, {\bf
17}:257-278.}

{ Nagumo J., Arimoto S. and Yoshizava S. [1962] An active pulse
transmission line simulating  nerve axon, Proc IREE {\bf 50}:
2061.}

{ Murray J.D. [1990] Mathematical Biology. Springer, New York.}

{ Buri\' c N. and Todorovi\'c D. [2003] Dynamics of
FitzHugh-Nagumo excitable systems with delayed coupling.
 Phys.Rev.E, {\bf 67}:066222.}

{  Meinardus G. and Nuernberg G. (eds.) [1985]  Delay Equations,
Approximation and Application. Birkhauser Verlag,
 Basel-Boston-Studgart.}

{ Faro J. and Velasco S. [1997] An approximation for prey-predator
 models with time delay. Physica D110: 313-322.}

{Arrowsimth D. [1990] An Introduction to Dynamical Systems.
Cambridge University Press, Cambridge}

{ Kuznetsov Y. [1995] Elements of Applied Bifurcation Theory.
Springer-Verlag, New York.}

{Coullet P. and Spiegel E. [1983] Amplitude equations for systems
with competing instabilities. SIAM J. Appl. Math. 776-821}

{ Kuznetsov Y. [1997] Explicit Normal Form Coefficients for all
Codim 2 Bifurcations of Equilibria in ODEs,  Report MAS-R9730.}

{ Ramana Reddy D.V., Sen A. and  Johnston G.L. [1999] Time delay
effects on
 coupled limit cycle oscillators at Hopf bifurcation.   Physica D, {\bf 129}: 15.}

{Faria T. and Magalhaes L.T. [1995] Normal Forms for Retarded
Functional Differential Equations with Parameters and Applications
to Hopf Bifurcation. J. Differential Equations, {\bf 122},
181-200.}

{ Shayer L.P. and Campbell S.A. [2000] Stability, Bifurcation, and
Multistability in a System of Two Coupled Neurons with Multiple
Time Delays. SIAM J. App. Math. {\bf 61}: 673-700.}

\newpage
{\bf FIGURE CAPTIONS}

 Figure 1: (a) Bifurcation set of the approximate system, the encircled area is enlarged in (b);
  (b) Dynamics near the codimension 2 generalized
  Hopf bifurcation.

 Figure 2: Bifurcation sets of the exact (thick lines) and the
 approximate (thin lines) systems. $\tau_1$ and $\tau_2$ are Hopf
 bifurcation curves of the exact system, and $\tau_{H,app}$ and
 $\tau_{p,app}$ are Hopf and pitchfork bifurcation lines of the
 approximate system.

 Figure 3:  Projections on $(x_1,x_2)$ of typical orbits approaching
  the possible attractors of the approximate
 system: a) One stable limit cycle (symmetric), b) stable stationary solution and the
stable limit cycle (symmetric), c)one stable stationary solution,
d) two stable stationary solutions.

 Figure 4: Projections on $(x_1,x_2)$ of typical orbits approaching
  the possible attractors of the exact
 system.  a) One stable limit cycle (symmetric), b) stable stationary solution and the
stable limit cycle (symmetric), c)one stable stationary solution,
d) one stable (asymmetric) limit cycle.

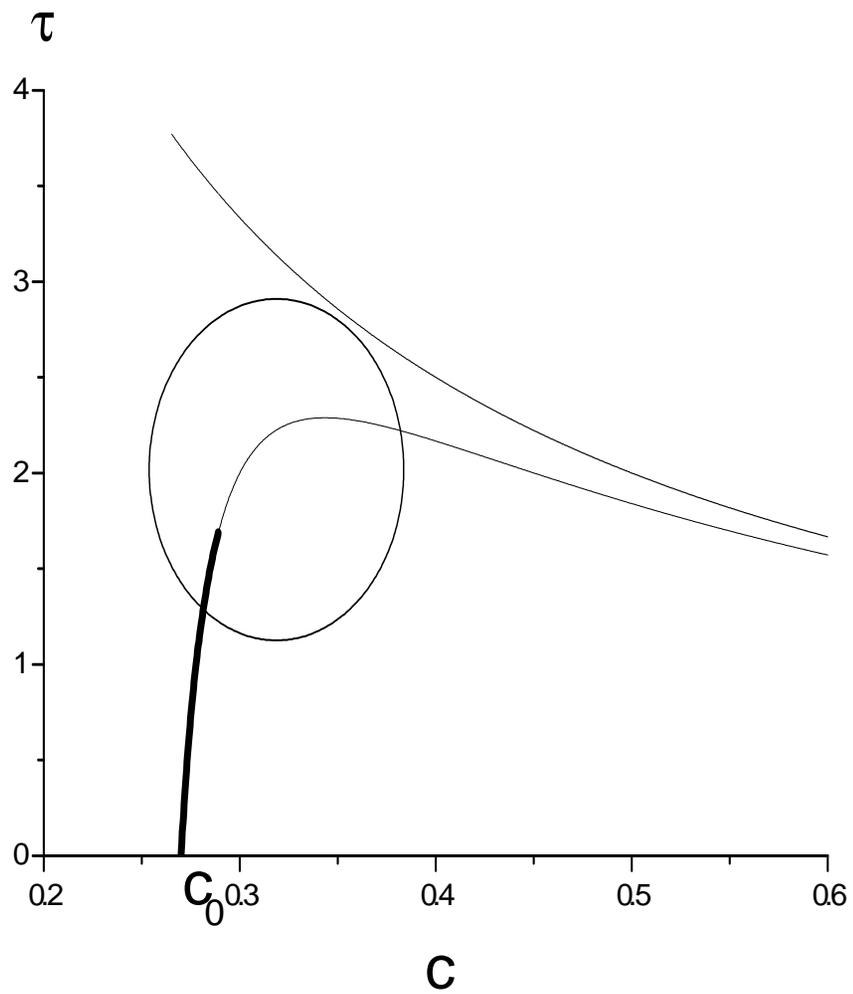
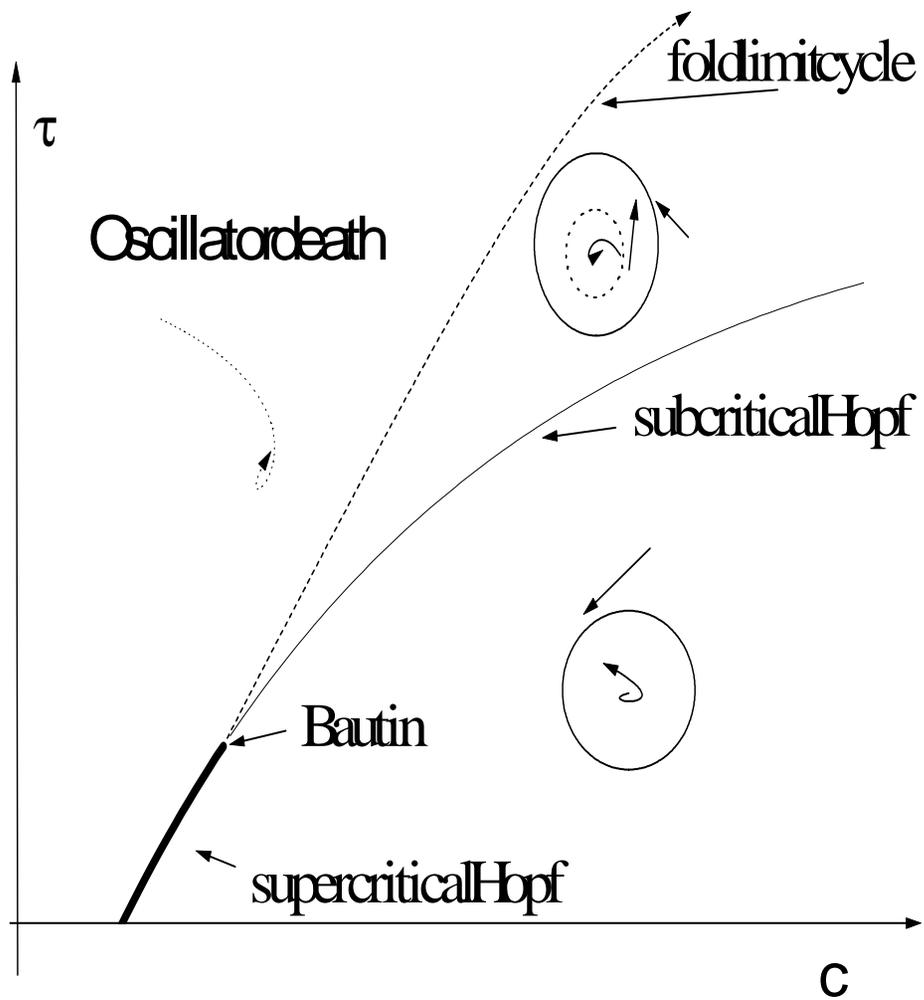
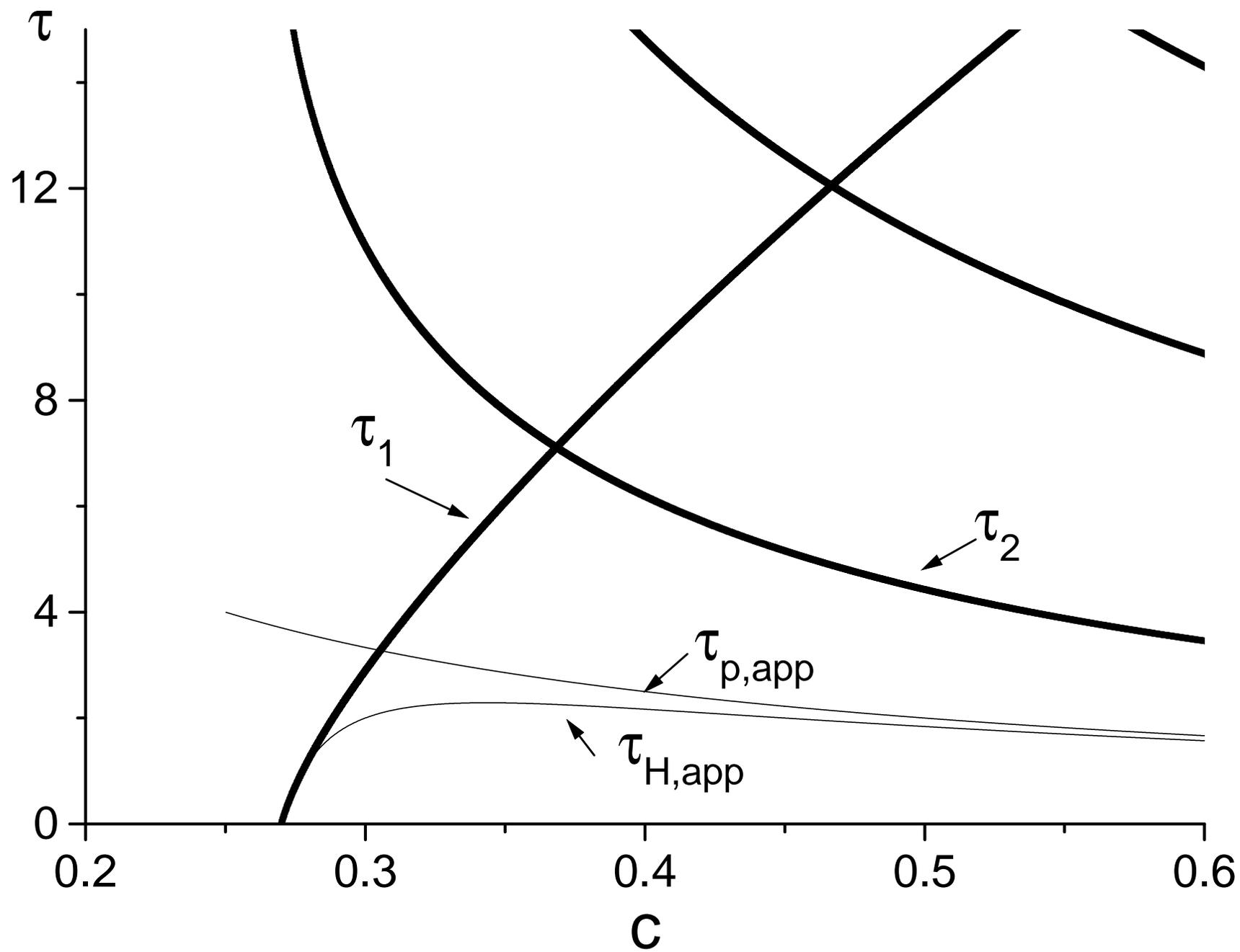
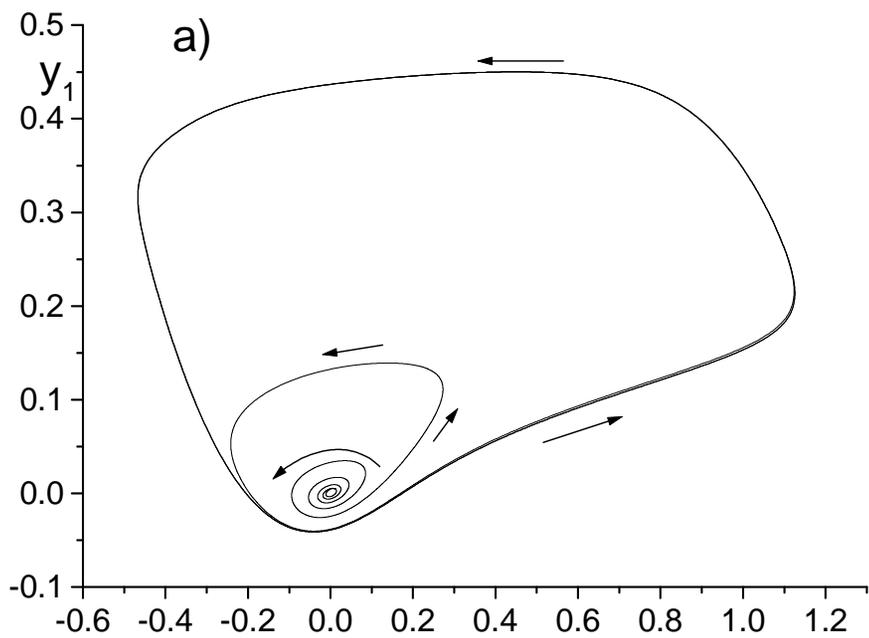
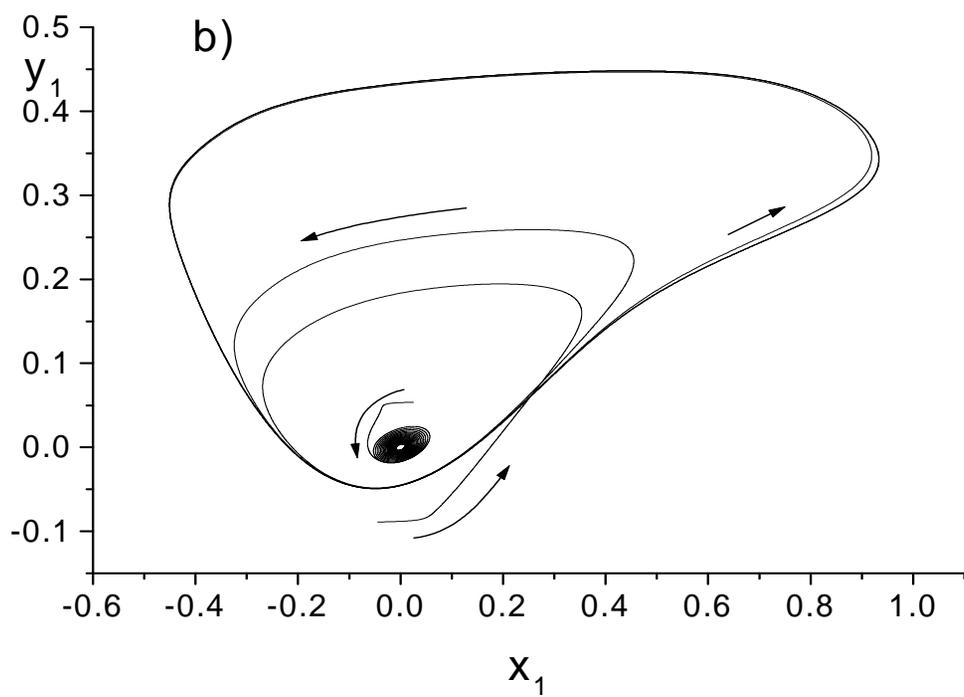
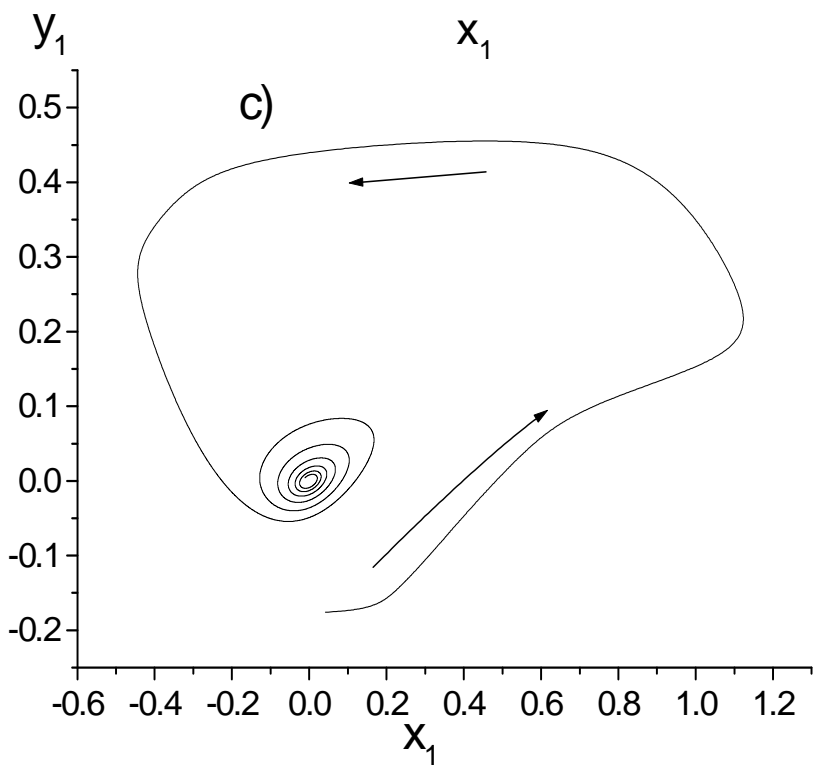
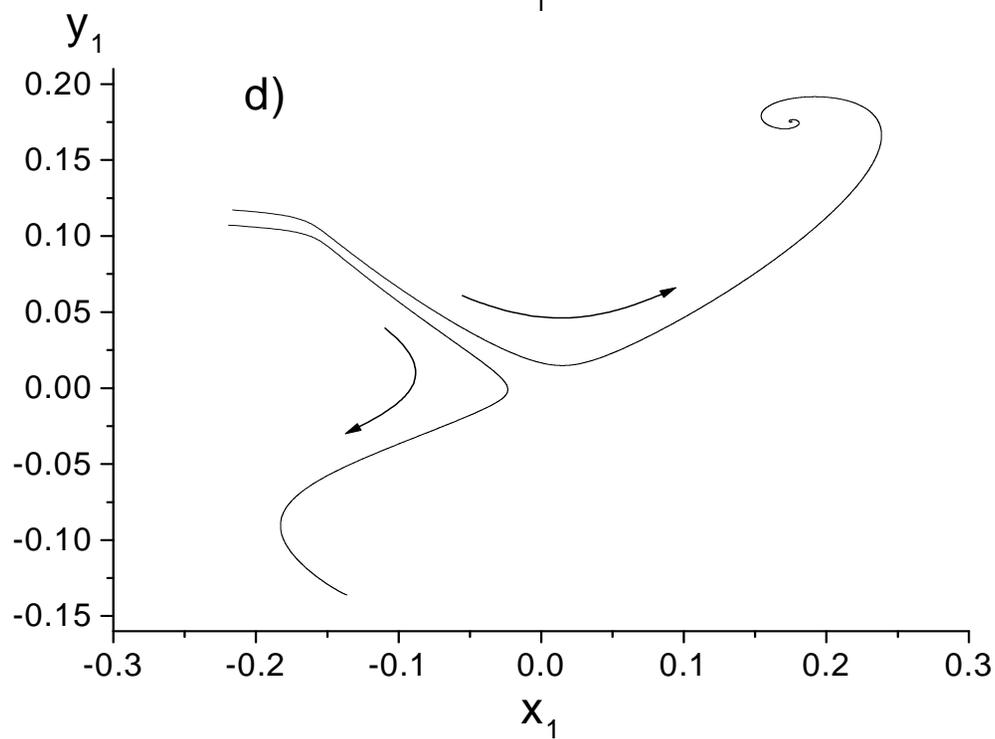
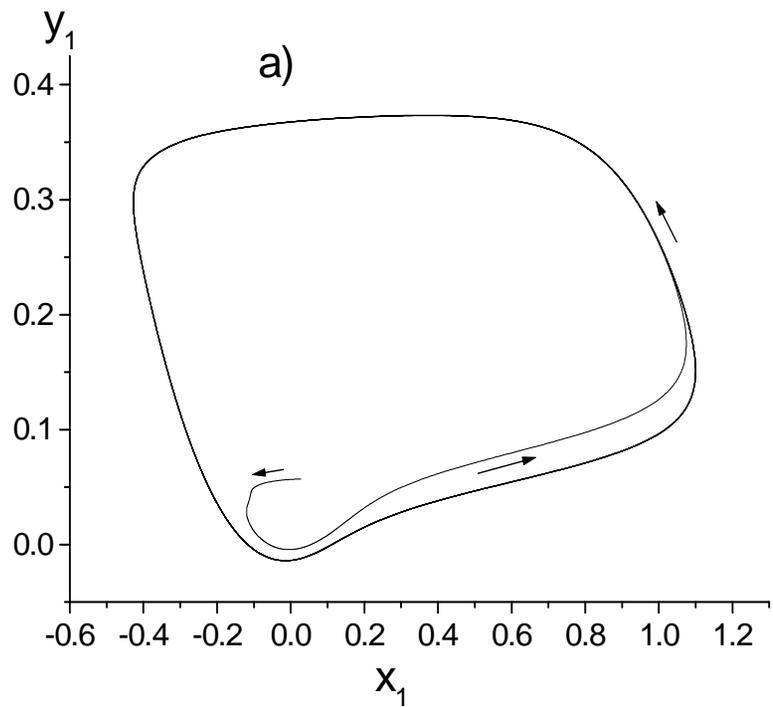
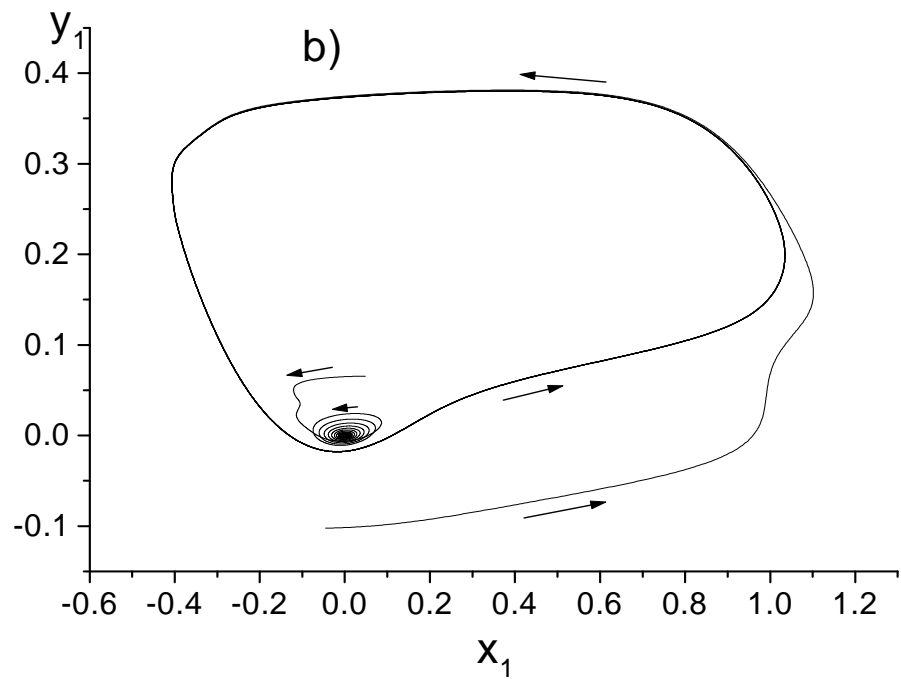
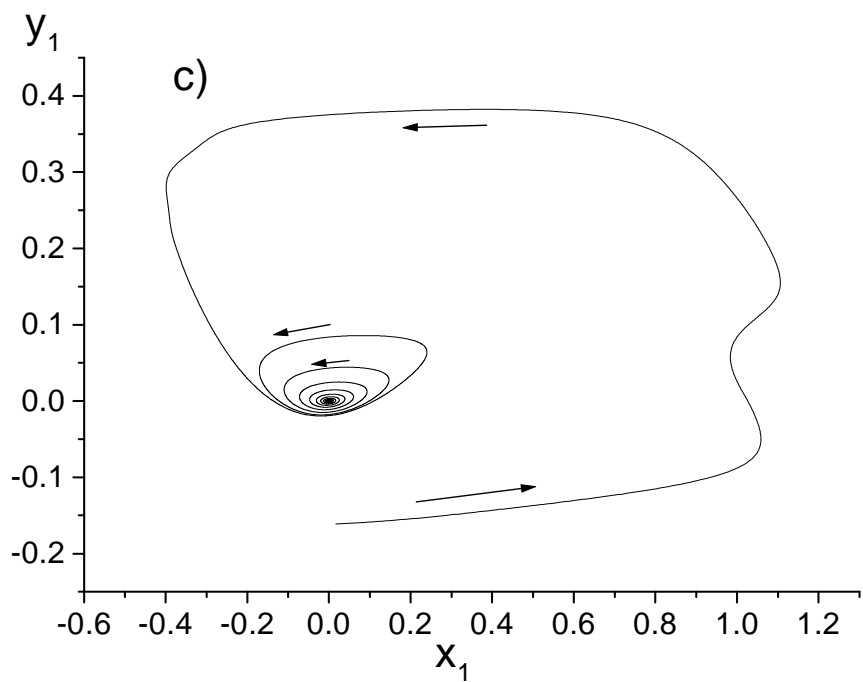
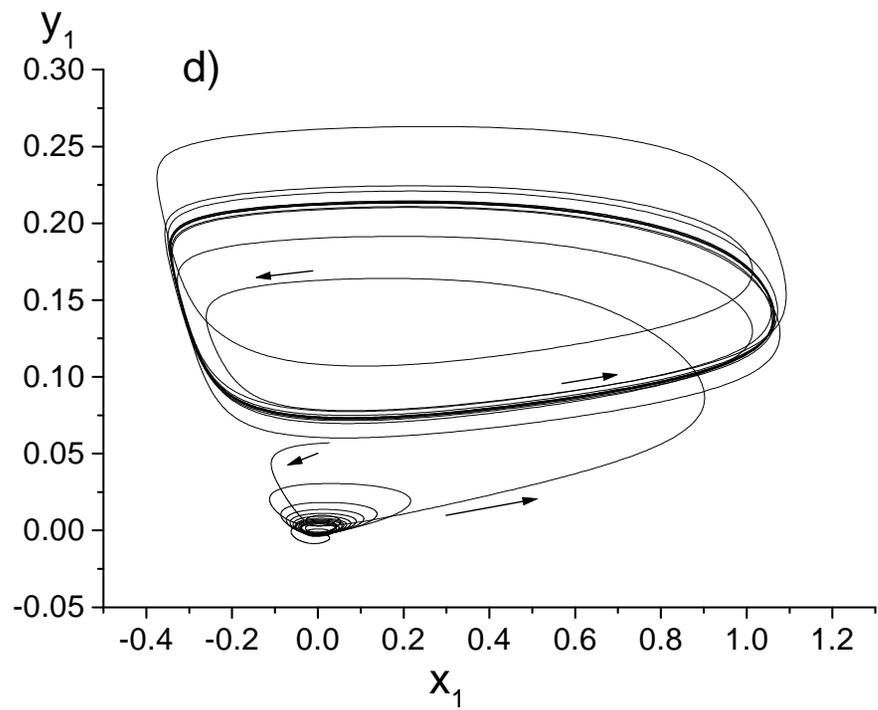
\end{document}